\documentclass[3p,times,twocolumn]{elsarticle}

\usepackage{ecrc}
\usepackage{amsmath}

\volume{00}

\firstpage{1}

\journalname{Nuclear Physics B Proceedings Supplement}

\runauth{A. Bussone et al.}

\jid{nuphbp}

\jnltitlelogo{Nuclear Physics B Proceedings Supplement}

\usepackage{amssymb}
\usepackage[figuresright]{rotating}

\begin{document}

\begin{frontmatter}



\dochead{}

\title{Heavy flavour precision physics from $N_f = 2 + 1 + 1$ lattice simulations}

%% use optional labels to link authors explicitly to addresses:F

\author[tov]{A.~Bussone} \author[infn-rm3]{N. Carrasco} \author[cf,tov]{P.~Dimopoulos\corref{cor1}} 
\author[tov]{R.~Frezzotti} \author[rm3]{P.~Lami} \author[rm3]{V. Lubicz} \author[tov]{F.~Nazzaro}
\author[rm3]{E.~Picca} \author[rm3]{L.~Riggio} \author[tov]{G.C.~Rossi} \author[sat]{F. Sanfilippo} \author[infn-rm3,rm3]{S.~Simula} 
\author[rm3]{C.~Tarantino}

\cortext[cor1]{Corresponding author, dimopoulos@roma2.infn.it}
%\fntext[fn1]{dimopoulos@roma2.infn.it}

\address[tov]{Dipartimento di Fisica, Universit\`a di Roma ``Tor Vergata", Via della Ricerca Scientifica 1, 00173, Rome, Italy}
\address[infn-rm3]{INFN, Sezione di Roma Tre, Via della Vasca Navale 84, I-00146 Rome, Italy}
\address[rm3]{Dipartimento di Fisica, Universit\`a  Roma Tre, Via della Vasca Navale 84, I-00146 Rome, Italy}
\address[cf]{Museo Storico della Fisica e Centro Studi e Ricerche ``Enrico Fermi", Compendio del Viminale,  
Piazza del Viminale 1, I-00184 Rome, Italy }
\address[sat]{School of Physics and Astronomy, University of Southampton, SO17 1BJ Southampton, U.K.}

\begin{abstract}
We present precision lattice calculations of the pseudoscalar decay constants of the charmed sector as well as determinations
of the bottom quark mass and its ratio to the charm quark mass. We employ $N_f=2+1+1$ dynamical quark gauge configurations 
generated by the European Twisted Mass Collaboration, using data at three values of the lattice spacing and pion masses as low as 210 MeV. 
Strange and charm sea quark masses are close to their physical values.
\end{abstract}


\end{frontmatter}

% \linenumbers

\section{Introduction}
\label{Intro}

Physical processes in the heavy quark sector offer the possibility to get some of the 
more stringest tests of the Standard Model and to search  
for possible footprints of New Physics dynamics, by directly challenging  
the unitarity constraints of the CKM matrix.

Lattice QCD  has already entered the
precision era as the accuracy of numerical computations is becoming comparable to that of 
experiments. 
For some of the relevant hadronic quantities in Flavour Physics 
the goal of per cent precision has been achieved. 
State-of-the-art lattice calculations involve O$(a)$-improved fermionic actions with $N_f=$ 2, 2+1 and 2+1+1 dynamical flavours 
with the smallest simulated pion masses being today at the physical point or slightly higher and employing three or more 
values of the lattice spacing. For a review with a critical evaluation of lattice results and averages, see~\cite{Aoki:2013ldr}. 
First computations with four non-degenerate quark flavours including electromagnetic effects 
have also been presented recently~\cite{Borsanyi:2014jba}. 
 
Direct computations by many lattice collaborations have shown that the cutoff effects in the $D$-sector are 
small and under control. Moreover, considerable progress has been recently made in flavour physics 
at the $b$ mass, with the help of both  effective theories approaches and thanks to  the
implementation of some innovative methods. 
All these progresses have allowed the determination of a number of $B$-physics parameters (e.g. $m_b$, $f_B$ and 
$f_{B_s}$) with controlled systematic uncertainties. 

Lattice methods are an invaluable tool to obtain direct determinations of hadronic 
quantities relevant for the computation of many of the  so called golden plated processes 
such as decay constants, form factors and bag parameters. 
For instance, the width of the $D$ and $D_s$ leptonic decays is given, to lowest order, 
 by 
\begin{equation}\label{eq:width}
\hspace*{-0.3cm}\Gamma(D_{(s)} \rightarrow \ell \nu) = \dfrac{G_F^2 f_{D_{(s)}}^2 m_{\ell}^2 M_{D{(s)}}}{8\pi}  
\left( 1 - \dfrac{m_{\ell}^2}{M_{D{(s)}}^2}\right)^2 
| V_{cd(s)}|^2 .  
\end{equation}      
Thus lattice computations of the quantities $f_D$ and $f_{Ds}$ gives access to the determination of the 
CKM matrix elements, $|V_{cd}|$ and $|V_{cs}|$, respectively, as in Eq.~(\ref{eq:width}) all the rest is known experimentally. 
On the experimental side also the accuracy of the measurements of the $D$~\cite{Zupanc:2013byn, delAmoSanchez:2010jg, Naik:2009tk}
and $D_s$~\cite{BESIII:2014uxa, BESIII:ichep2014} leptonic width has 
progressively improved during the years.

Lattice QCD provides a first principles' way to compute quark masses. This is possible since quark masses 
enter as parameters in the QCD Lagrangian and their values can be extracted by matching hadron masses 
calculated on the lattice with their experimental values. The accuracy of quark mass estimates depends  
on the conversion from the lattice regularisation to continuum renormalisation schemes. Quark mass ratios instead  
can be computed in a fully non-perturbative way and are free of renormalisation scheme ambiguities. 
We notice, here, that the knowledge of the $b$-quark mass value and to less extent that of the $c$-quark mass plays an important 
r$\hat{{\rm o}}$le in the study of the Higgs decay to $b\bar b$ and $c\bar c$~\cite{Djouadi:2005gi}. 
 
The European Twisted Mass Collaboration (ETMC) has undertaken an extensive program of heavy quark physics calculations on the lattice 
using two and four dynamical flavours. Here we present the results of the computation of the $D_{(s)}$ pseudoscalar meson decay 
constants (in the isospin symmetric limit) and the $b$ to $c$ quark mass ratio 
obtained using  gauge configurations with $N_f=2+1+1$ dynamical quarks. The main ({\it preliminary}) results  
in these proceedings are 
\begin{equation}
f_D = 208.7(5.2) ~{\rm MeV}, ~~~f_{Ds} = 247.5(4.1) ~ {\rm MeV},
\end{equation} 
\begin{equation}
\dfrac{f_{Ds}}{f_D} = 1.186(21),~~~~~~ \left(\dfrac{f_{Ds}}{f_D}\right) \Big{/} \left(\dfrac{f_{K}}{f_{\pi}}\right)=0.998(14),
\end{equation}
\begin{equation}
m_b(\overline{{\rm MS}}, m_b) = 4.26 (16) ~ {\rm GeV},  
\end{equation}
\begin{equation}
m_b/m_c = 4.40(8)
\end{equation}
For completeness we remind the recent ETMC determinations of the $c$-quark mass and the 
charm to strange quark mass ratio published in~\cite{Carrasco:2014cwa}: 
\begin{equation}
m_c(\overline{{\rm MS}}, m_c) = 1.348 (42) ~ {\rm GeV}, ~~ m_c/m_s = 11.62(16)
\end{equation}
For a preliminary ETMC computation of the $B$-meson decay constants, giving 
$f_B = 196 (9)$ MeV, $f_{Bs} = 235 (9)$ MeV and $f_{Bs}/f_{B} = 1.201 (25)$,  
we refer to~\cite{Carrasco:2013naa}.

\section{Lattice setup}
%\label{Latsetup}
ETMC has produced gauge configurations with $N_f=2+1+1$ dynamical quarks~\cite{Baron:2010bv} employing the Iwasaki gluon 
action~\cite{Iwasaki:1985we} 
and the Wilson Twisted Mass fermionic action for the sea quarks~\cite{Frezzotti:2003xj}. Automatic O$(a)$-improvement is guaranteed 
both for the light and heavier quarks by tuning at maximal twist whilst 
the drawback of the mixing between the strange and charm sectors~\cite{Baron:2010th} is avoided in the valence by 
using the Osterwalder-Seiler fermions~\cite{Osterwalder:1977pc}. We have data ensembles 
at three values of the lattice spacing 
in the range [0.06, 0.09] fm. Simulated pion masses lie in the interval [210, 440] MeV. 
Thanks to the properties of the twisted mass action light quarks in the sea and all types of quarks in the valence 
enjoy multiplicative mass renormalisation, $Z_m = 1/Z_P$, which is computed non-perturbatively using the RI$'$-MOM 
scheme~\cite{Carrasco:2014cwa}. 
Moreover owing to PCAC, at maximal twisted angle no normalisation constant is needed in the computation of the decay constants.
In Ref.~\cite{Carrasco:2014cwa} we have presented our computation for the quark masses of 
the (degenerate) light $m_{u/d}(\overline{{\rm MS}}, 2~{\rm GeV})= 3.70 (17)$ MeV, strange 
$m_{s}(\overline{{\rm MS}}, 2~{\rm GeV})= 99.6 (4.1)$ MeV and  charm $m_{c}(\overline{{\rm MS}}, m_c)= 1.348 (42)$ GeV, which 
are determined by using the experimental values of the pion, kaon and  
$D$ (or $D_s$) masses, respectively. The phenomenological value of $f_{\pi}$ has been used for setting the scale. \\
In this work the computation of the decay constants in the charmed region 
as well as the determination of the $b$-quark mass are performed using (Gaussian) smearing meson 
operators~\cite{Gusken:1989qx, Jansen:2008si} combined with APE smeared links~\cite{Albanese:1987ds}
in order to reduce both the coupling of the interpolating field with the excited states and the gauge noise 
of the links involved in the smeared fields. (For an alternative preliminary analysis of the charmed decay constants that use 
local point correlators see Ref.\cite{Dimopoulos:2013qfa}.) 
A  summary of the most important details of our simulations is given in Table~1.         

\begin{table}[!h] \label{Tab1}
\begin{center}
%\footnotesize
\scalebox{0.64}{
\begin{tabular}{||c|c|c|c|c|c||}
\hline
 $\beta$ & $V / a^4$ &$a\mu_{sea}=a\mu_\ell$ & $N_{cfg}$& $a\mu_s$& $a\mu_c - a\mu_h$ \\
\hline
 $1.90$ & $32^{3}\times 64$ &$0.0030$ &$150$ &  $0.0180,$& $0.21256, 0.25000,$ \\
        & & $0.0040$ &  $150$ &$0.0220,$ & 0.29404, 0.34583, \\
        & & $0.0050$ &   $150$ &$ 0.0260$ &  0.40675, 0.47840, \\
        & &          &         &          &  0.56267, 0.66178, \\
        & &          &         &          &  0.77836,  \\  %0.91546
\cline{1-4}
 $1.90$ & $24^{3}\times 48 $ & $0.0040$ & $150$ & & \\
        & & $0.0060$ &   $150$ &&  \\
        & & $0.0080$ &   $150$ & &  \\
        & & $0.0100$ &   $150$ & &  \\
\hline
 $1.95$ & $32^{3}\times 64$ &$0.0025$ & $150$& $0.0155,$& $0.18705, 0.22000,$ \\
        & & $0.0035$ &  $150$ &$  0.0190,$ &0.25875, 0.30433,\\
        & & $0.0055$ &  $150$ &$ 0.0225$    &0.35794, 0.42099, \\
        & & $0.0075$ &  $150$  &             &0.49515, 0.58237   \\
        & &          &        &             &0.68495,  \\  %0.80561
\cline{1-4}
 $1.95$ & $24^{3}\times 48 $ & $0.0085$  & $150$ & & \\
\hline
 $2.10$ & $48^{3}\times 96$ &$0.0015$ & $90$& $0.0123,$& $0.14454, 0.17000, $ \\
        & & $0.0020$ &   $90$ &$0.0150,$ &0.19995, 0.23517, \\
        & & $0.0030$ &  $90$ & $  0.0177$  &0.27659, 0.32531,  \\
        & &          &       &             &0.38262, 0.45001, \\
        & &          &       &             &0.52928,  \\ %0.62252
 \hline
\end{tabular}
}
\vspace*{-0.15cm}
\caption{Summary of simulation details. Gauge couplings $\beta$ = 1.90, 1.95 and 2.10 correspond to lattice spacings $a$ = 0.089, 0.082 and 
0.062, respectively; see Ref.~\cite{Carrasco:2014cwa}. We denote with $a\mu_{\ell}$, $a\mu_{s}$ and $a\mu_{c}-a\mu_{h}$, the light, 
strange-like, charm-like and somewhat heavier bare quark masses, respectively, entering in the valence sector computations. } 
\end{center}
\end{table} 

\section{Charmed decay constants}
We use two point correlation functions with pseudoscalar interpolating operators, 
$P(x) = \overline{q}_1(x) \gamma_5 q_2(x)$, that in periodic lattice have the typical form: 
\begin{eqnarray}
C_{PP}(t) &=& (1/L^3)\sum_{\vec{x}} \langle P(\vec{x},t) P^{\dagger}(\vec{0}, 0) \rangle \nonumber \\
          && \hspace*{-1.2cm}\stackrel{t \gg 0, ~(T-t) \gg 0}{\longrightarrow}
          \dfrac{\xi_{PP}}{2M_{ps}} \left(e^{-M_{ps} t} + e^{-M_{ps}(T-t)} \right)
\end{eqnarray}          
We take the Wilson parameters of the two valence quarks of the pseudoscalar meson to be opposite in order 
to guarantee that the cutoff effects 
on the pseudoscalar mass are 
$O(a^2 \mu)$~\cite{FrezzoRoss1, Frezzotti:2005gi, Dimopoulos:2009qv}. 
We then consider two cases, using smeared source only and   
source and sink both smeared. As for $\xi_{PP}$, this is given by  
$\xi_{PP} = \langle 0| P^{L}|ps \rangle \langle ps |P^{S}|0 \rangle$ in the first case and 
$\xi_{PP} = \langle 0 | P^{S}|ps \rangle\langle ps |P^{S}|0\rangle$ in the second one,  
where $L$ and $S$ indicate local and smeared operators.   
By combining the two kinds of correlators it is easy to obtain the matrix element of the local operator  
$g_{ps} = \langle 0 | P^{L} | ps \rangle $ which serves for computing the pseudoscalar decay constant (via PCAC) 
given by:
\begin{equation} \label{eq:fps}
f_{ps} = (\mu_{1} + \mu_{2}) \dfrac{g_{ps}}{M_{ps} \sinh M_{ps}}, 
\end{equation} 
where $\mu_{1,2}$  are the masses of the valence quarks that form the pseudoscalar meson with mass $M_{ps}$. 
The use of $\sinh M_{ps}$ (rather than $M_{ps}$) in Eq.~(\ref{eq:fps}) 
is beneficial for reducing the discretisation errors. 
For the computation of $f_{Ds}$ we tune, via well controlled interpolations, 
one of the valence quark masses to the value of the strange  mass 
and the other to the value of the charm  mass, both taken from Ref.~\cite{Carrasco:2014cwa}. 
\begin{figure}[!t]
\hspace*{-1.2cm} {\includegraphics[scale=0.58,angle=-0]{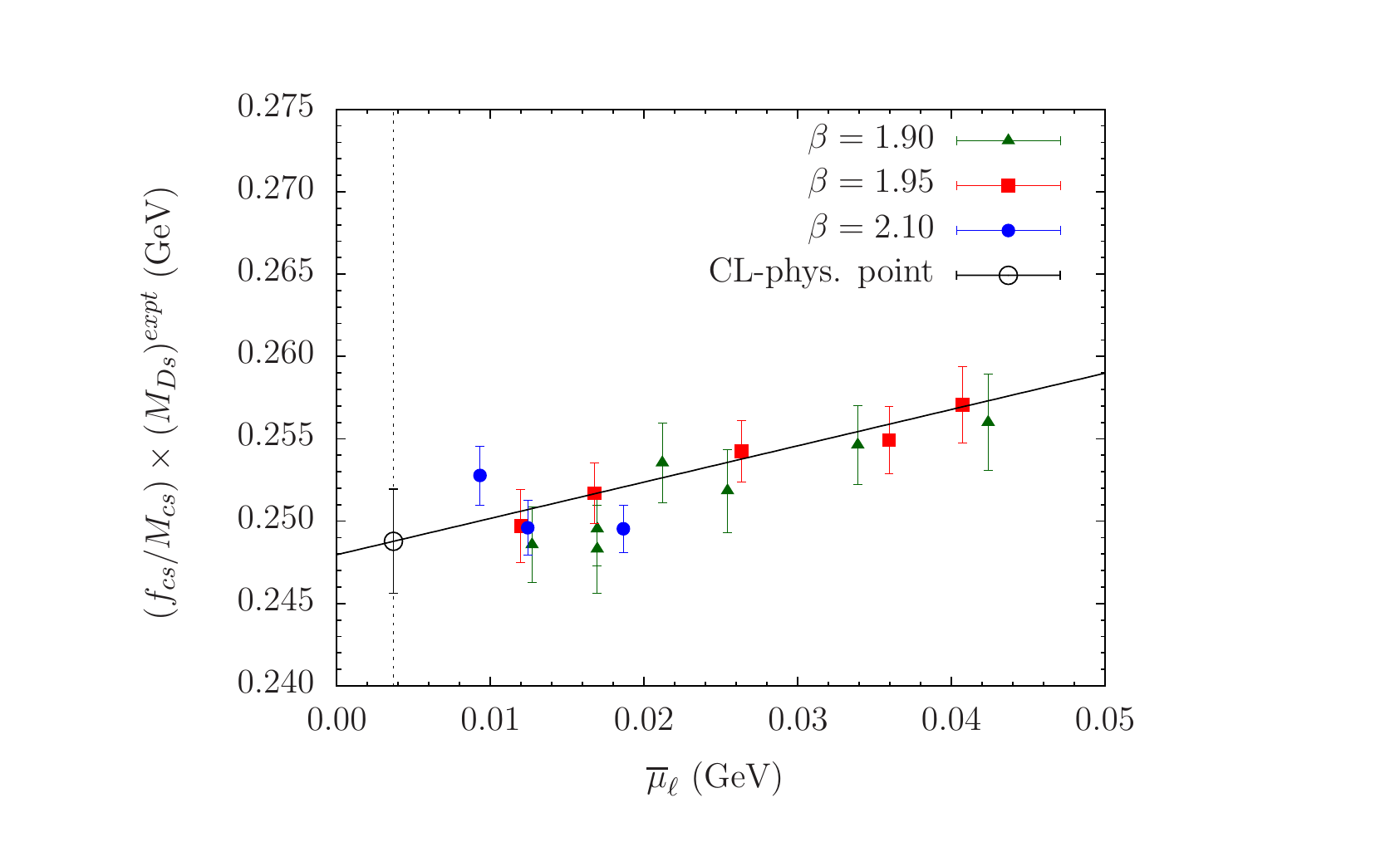}}
\vspace*{-0.1cm}
\caption{ Combined chiral and continuum fit ($\chi^2/(dof) = 0.8$) of $(f_{cs}/M_{cs}) \times M_{Ds}^{expt}$ against the
renormalised light quark mass expressed in $\overline{{\rm MS}}$-scheme at the scale of 2 GeV, $\overline\mu_{\ell} = 
\overline{\mu}_{sea}$. 
The fit ansatz is linear both in $\overline\mu_{\ell}$ and in $a^2$. 
The vertical black thin line marks the position of $u/d$ quark mass point.
The empty black circle is our result at the physical $u/d$ quark mass point in the continuum
limit.
  }
\label{fig:fDs}
\end{figure}
In this way, for each value of the sea light quark mass and of the three lattice spacings, we get estimates
for the decay constant $f_{cs}$. 
Then a simultaneous extrapolation to the physical value of the $u/d$ quark mass and to the continuum limit 
can be performed in order to obtain $f_{Ds}$. In the present analysis we consider the quantity 
$(f_{cs}/M_{cs}) \times M_{Ds}^{expt}$, where $M_{cs}$ is a pseudoscalar meson mass  
made of $c$ and $s$ quarks and is computed at each value of the 
sea light quark mass,  while $M_{Ds}^{expt} = 1969.0(1.4)$ MeV is the experimental value of the $D_s$ mass. 
The above choice of observable is advantageous 
because, first, in the determination of $f_{Ds}$ any scale setting uncertainty is avoided and, second, this quantity 
presents very small discretisation effects. The fit ansatz of the combined chiral and continuum extrapolation reads: 
$[(f_{cs}/M_{cs}) \times M_{Ds}^{expt}] = C_0 + C_1~ \overline{\mu}_{\ell} + D~ a^2$, see Fig.~\ref{fig:fDs}. We control chiral 
fit uncertainties by adding in the above fit ansatz either a quadratic quark mass term  or fitting data corresponding to 
light pseudoscalar masses with $M_{\ell \ell} < 350$ MeV. Finite volume systematics are estimated 
by fitting data corresponding to  $L > 2.6$ fm. 
Discretisation systematic errors have been estimated by fitting data either from the two finest lattice spacings or
from the two coarsest ones, and also by estimating the difference of our results from the finest lattice to the continuum limit.
Moreover, we have also included the propagated error due to the $m_{s,c}$ uncertainties as well as 
the systematic effect of the quark mass renormalisation constant (RC) 
computed in two ways that differ  by O$(a^2)$ effects. 
Our central value is the weighted average over the results from 
all the analyses described above. Our ({\it preliminary}) result  for $f_{Ds}$ reads
\begin{equation}\label{eq:fDs_res}
f_{Ds} = 247.5\,(3.0)_{stat+fit}\, (2.7)_{syst}\,[4.1] ~~{\rm MeV},   
\end{equation}
where we report in square brackets the total error ($\sim$ 1.6\%) that is the sum in quadrature of the statistical 
and systematic uncertainties. 
For the full error budget see Table~2.  

\begin{table}[!h] \label{Tab: budget}
\begin{center}
%\footnotesize
\scalebox{0.85}{
\begin{tabular}{|l|c|c|c|}
\hline
uncertainty (in \%)  & $f_{Ds}$ & $f_{Ds}/f_{D}$ & $f_{D}$\tabularnewline
\hline
\hline
stat. + fit & 1.2 & 0.8 & 1.6\tabularnewline
\hline
syst. from chiral fits & 0.8 & 0.8 & 1.1\tabularnewline
\hline
syst. from discr. effects & 0.8 & 0.7 & 1.0\tabularnewline
\hline
syst. from FSE & 0.1 & 0.4 & 0.4\tabularnewline
\hline
syst. from $f_K/f_{\pi}$ & - & 1.2 & 1.2\tabularnewline
\hline
Total & 1.6 & 1.8 & 2.5\tabularnewline
\hline
\end{tabular}
}
\vspace*{-0.15cm}
\caption{Full error budget for $f_{Ds}$, $f_{Ds}/f_D$ and $f_D$. The different sources of uncertainty are self explanatory.} 
\end{center}
\end{table}
          
\begin{figure}[!h]
\begin{center}
{\includegraphics[scale=0.60,angle=-0]{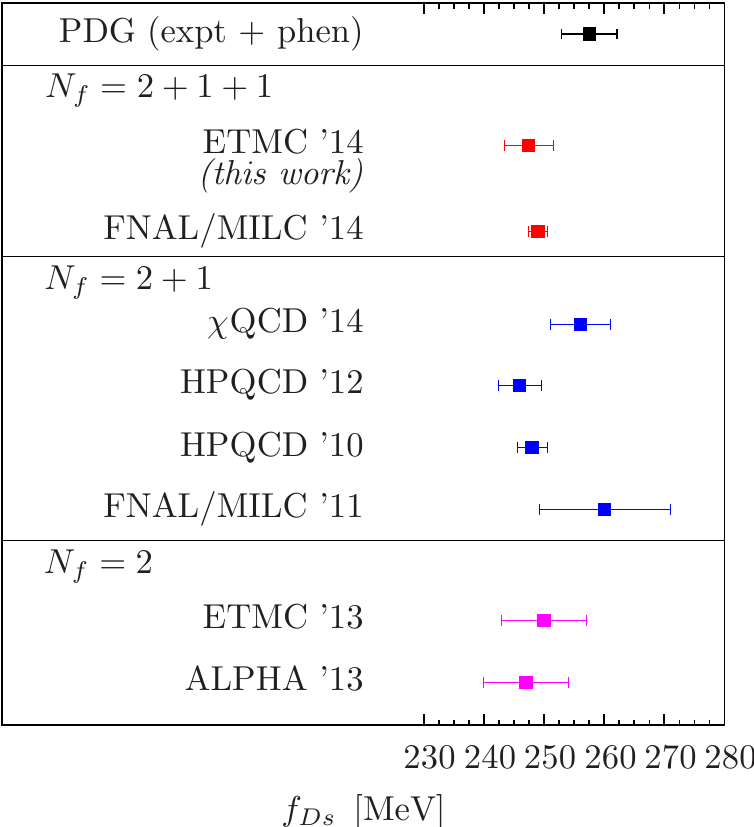}}
\vspace*{-0.1cm}
\caption{ We compare the available continuum limit determinations for $f_{Ds}$ (in MeV) from lattice studies that use 
$N_f=2$, 2+1 and 2+1+1 dynamical flavours. "ETMC '14" result refers to the present work. 
For the results of other lattice studies we refer to (from top to bottom)~\cite{Bazavov:2014wgs, Yang:2014sea, 
Na:2012iu, Davies:2010ip, Bazavov:2011aa, Heitger:2013oaa, Carrasco:2013zta}). For the PDG 
result see~\cite{Agashe:2014kda}. 
  }
\label{fig:fDs_compr}
\end{center}
\end{figure}

In Fig.~\ref{fig:fDs_compr}  we compare our result with those computed in other lattice studies and with the PDG estimate based 
on experimental results and unitarity assumptions. Some tension between the PDG estimate and 
the most precise lattice results is still present.  

\begin{figure}           
\hspace*{-1.5cm} {\includegraphics[scale=0.58,angle=-0]{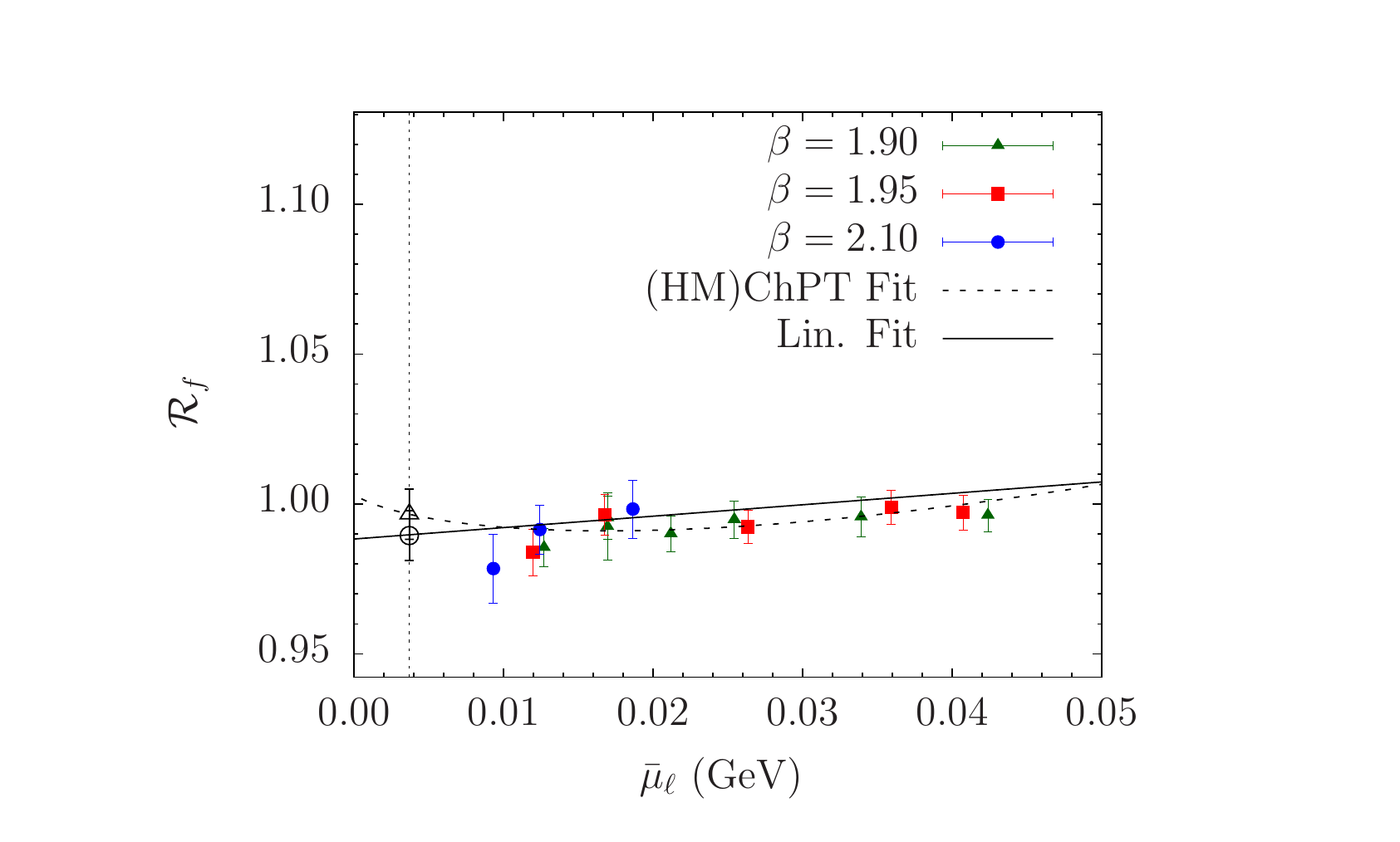}}
\caption{Combined chiral and continuum fit for the quantity 
${\cal R}_f$ against the
renormalised light quark mass expressed in $\overline{{\rm MS}}$-scheme at the scale of 2 GeV, $\overline\mu_{\ell} = 
\overline{\mu}_{sea}$. 
The vertical black thin line marks the position of $u/d$ quark mass point.
The empty black circle and empty triangle represent the results for the ratio $f_{Ds}/f_{D}$, 
using the fit ansatz of Eq.~(\ref{eq:fa_1}) ($\chi^2/(dof)=0.7$) and Eq.~(\ref{eq:fa_2}) ($\chi^2/(dof)=1.1$), respectively,  
at the physical $u/d$ quark mass point and in the continuum limit.
  }
\label{fig:fDs_ov_fD}
\end{figure}

In order to determine the SU(3) symmetry breaking ratio $f_{Ds}/f_D$ we measure on our data sets the double ratio 
${\cal R}_f=(f_{cs}/f_{c\ell})(f_{\ell\ell}/f_{s\ell})$. 
This choice enjoys the property of very mild light quark mass dependence as expected from the 
large cancellation between the SU(2) chiral logarithms~\cite{Becirevic:2002mh, Blossier:2009bx}. The quantity ${\cal R}_f$ 
in the continuum limit and at the physical pion mass point multiplied with  $(f_K/f_{\pi})$ 
(taken from Ref.~\cite{Carrasco_new})  will provide   
the result for $f_{Ds}/f_D$.  
We try the following fit ans\"atze:
\begin{eqnarray} \label{eq:fa_1}
{\cal R}_f &=& c_0^{(1)} + c_1^{(1)} \overline\mu_{\ell} + D^{(1)} a^2,  \\ 
{\cal R}_f &=& c_0^{(2)}\Big[1 + c_1^{(2)} \overline\mu_{\ell} + \nonumber \\ 
      &&      +\left( \frac{9\hat{g}^2}{4} - \frac{1}{2} \right) \label{eq:fa_2}
\xi_{\ell}\log\xi_{\ell}   \Big]  + D^{(2)} a^2, 
\end{eqnarray}
where  $\xi_{\ell} = (2 B_0 \overline\mu_{\ell})/(4 \pi f_0)^2$ with $B_0$ and $f_0$ determined in Ref.~\cite{Carrasco:2014cwa}. 
We have applied finite size corrections using Ref.\cite{Colangelo:2005gd}. Among  
the available estimates for the ($D^{*}D\pi$) coupling we have used $\hat{g}=0.61(7)$ that in our case leads to the most conservative  
estimate for the chiral fit systematic uncertainty. The chiral and continuum limit extrapolation is shown in Fig.~\ref{fig:fDs_ov_fD}. 
Moreover we have performed an analysis similar to the one for $f_{Ds}$ in order to estimate our systematic uncertainties. 
The full error budget is given in Table~2.  
The central value corresponds to the  weighted average over results from all the different analyses.    
Our ({\it preliminary})  results  read
\begin{equation}
(f_{Ds}/f_D){\large/}(f_{K}/f_{\pi}) = 0.998\,(8)_{stat+fit}\,(11)_{syst} [14], 
\end{equation}
\begin{equation}
f_{Ds}/f_{D} = 1.186\,(9)_{stat+fit}\,(19)_{syst}\,[21], \label{eq:fDs_ov_fD_res}
\end{equation}  
and each one of the total errors (in square brackets) is the sum in quadrature of 
the statistical error and the systematic one. 

We combine the results from Eqs.~(\ref{eq:fDs_res}) and (\ref{eq:fDs_ov_fD_res}) to get our ({\it preliminary}) result  
for the decay constant of the $D$-meson, namely $f_D = f_{Ds}/(f_{Ds}/f_D)$, which reads:
\begin{equation}\label{eq:fD_res}
f_D = 208.7\,(3.3)_{stat+fit}\,(4.0)_{syst}\, [5.2] ~~{\rm MeV} , 
\end{equation}
where also in this case the total error written in square brackets ($\sim$ 2.5\%) is the sum in quadrature 
of the statistical and systematic uncertainties. 
For the complete error budget see Table~2.

\begin{figure}
\begin{center}
{\includegraphics[scale=0.55,angle=-0]{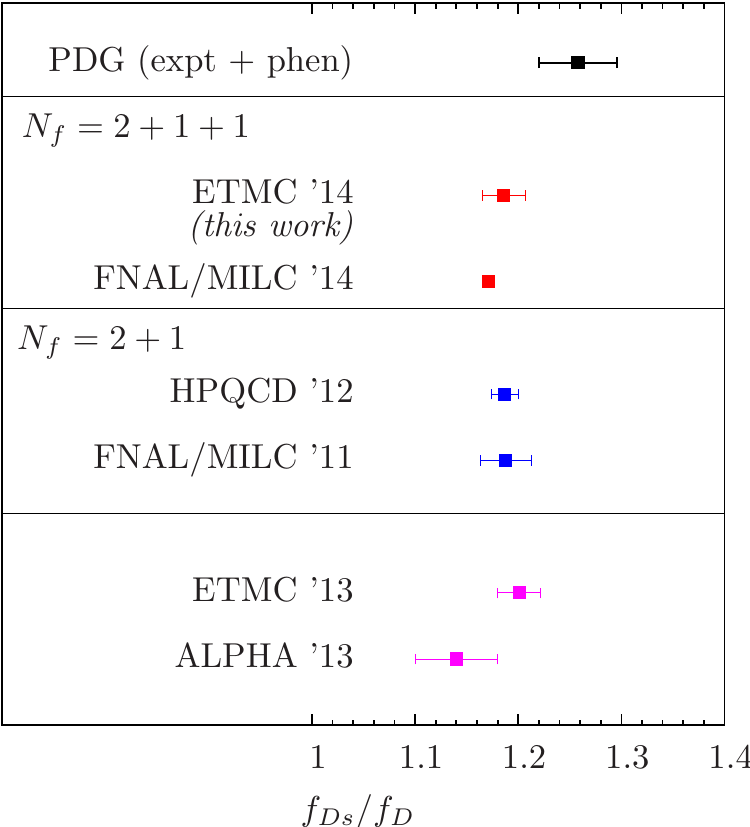}}
\caption{We compare the available continuum limit determinations for $f_{Ds}/f_D$ from lattice studies that use 
$N_f=2$, 2+1 and 2+1+1 dynamical flavours. "ETMC '14" result refers to the present work. 
For the results of the other lattice studies we refer to (from top to bottom)~\cite{Bazavov:2014wgs, 
Na:2012iu, Bazavov:2011aa, Heitger:2013oaa, Carrasco:2013zta}). For the PDG 
result see~\cite{Agashe:2014kda}. 
  }
\label{fig:fDs_ov_fD_compr}
\end{center}
\end{figure}

\begin{figure}[!t]
\begin{center}
{\includegraphics[scale=0.55,angle=-0]{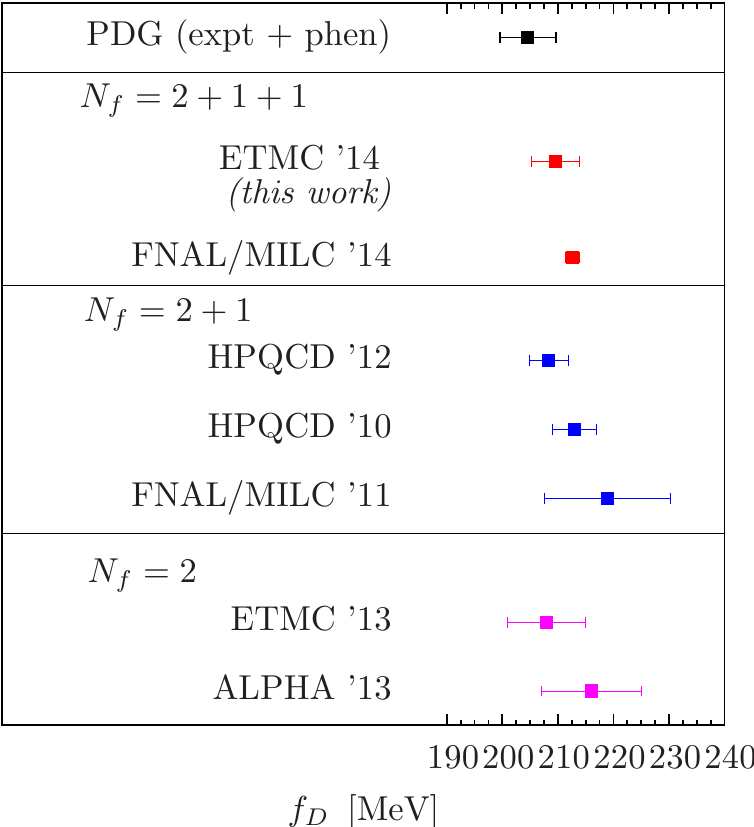}}
\caption{ We compare the available continuum limit determinations for $f_{D}$ (in MeV) from lattice studies that employ 
$N_f=2$, 2+1 and 2+1+1 dynamical flavours. "ETMC '14" result refers to the present work. 
For the results of the other lattice studies we refer to (from top to bottom)~\cite{Bazavov:2014wgs, Yang:2014sea, 
Na:2012iu, Davies:2010ip, Bazavov:2011aa, Heitger:2013oaa, Carrasco:2013zta}). For the PDG 
result see~\cite{Agashe:2014kda}. 
  }
\label{fig:fD_compr}
\end{center}
\end{figure}

In Figs.~\ref{fig:fDs_ov_fD_compr} and \ref{fig:fD_compr} we present a world comparison between lattice 
results for $f_{Ds}/f_{D}$ and $f_D$, respectively. 
In both figures the PDG estimate is also included. For some recent non-lattice estimates of the charmed decay constants, 
see Refs.~\cite{Narison:2012xy, Lucha:2011zp, Wang:2013ff, Gelhausen:2013wia}.

\section{Computation of $m_b$ and $m_b/m_c$}
\begin{figure}          
\hspace*{-1.0cm} {\includegraphics[scale=0.54,angle=-0]{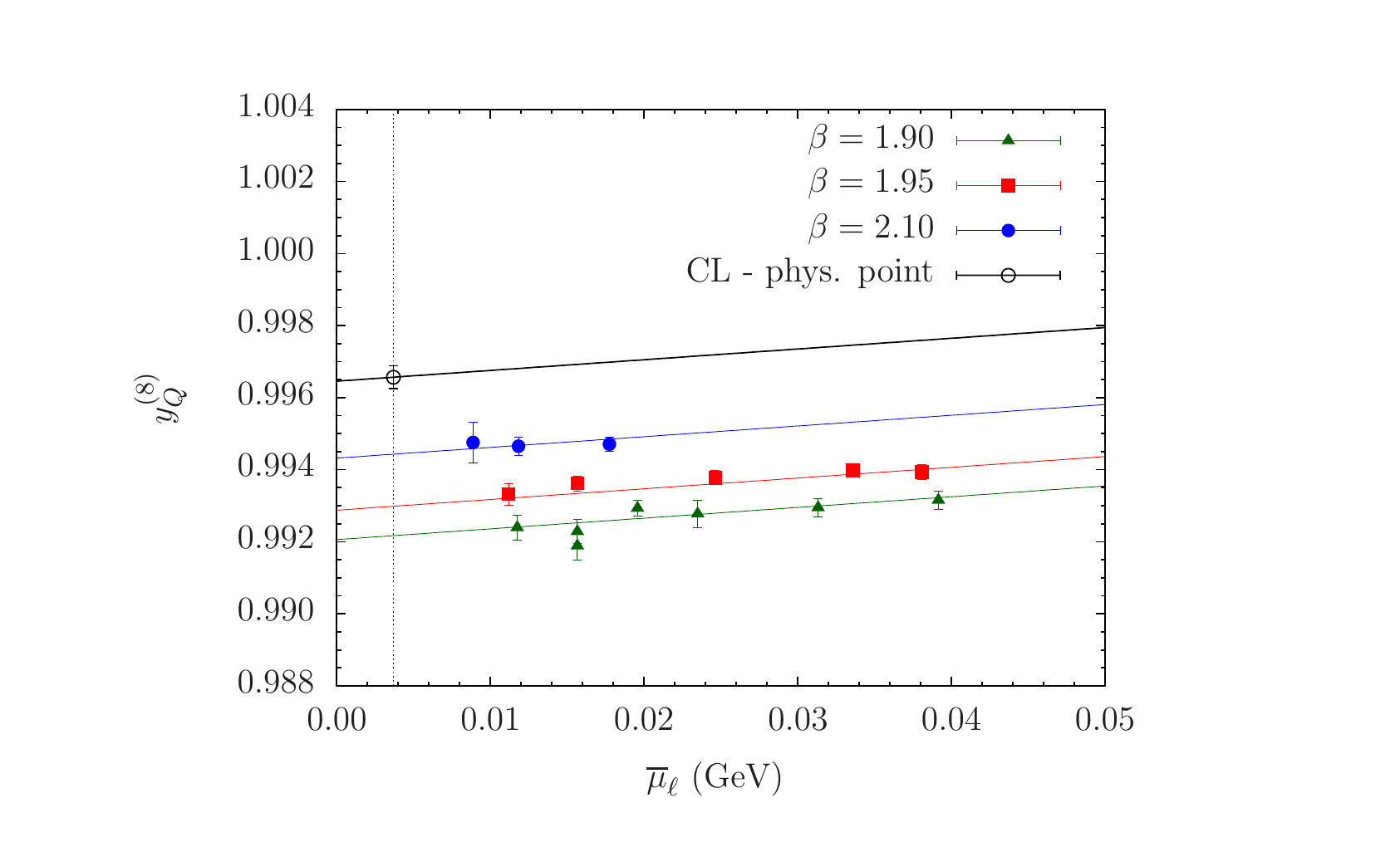}}
\caption{ Combined chiral and continuum fit of the ratio defined in Eq.~(\ref{eq:yn}) and corresponding to the case 
of the two largest values of heavy quark mass against the renormalised light
quark mass $\overline{\mu}_{\ell} = \overline{\mu}_{sea}$. The fit ansatz is linear both in $\overline{\mu}_{\ell}$
and in $a^2$ with $\chi/(dof) = 1.1$.  
The empty black circle is our result at the physical $u/d$ quark mass point in the continuum limit. In this example 
we have considered $\gamma = 0.75 $. 
  }
\label{fig:y8}
\end{figure}

\begin{figure}[!t]           
\hspace*{-1.0cm} {\includegraphics[scale=0.56,angle=-0]{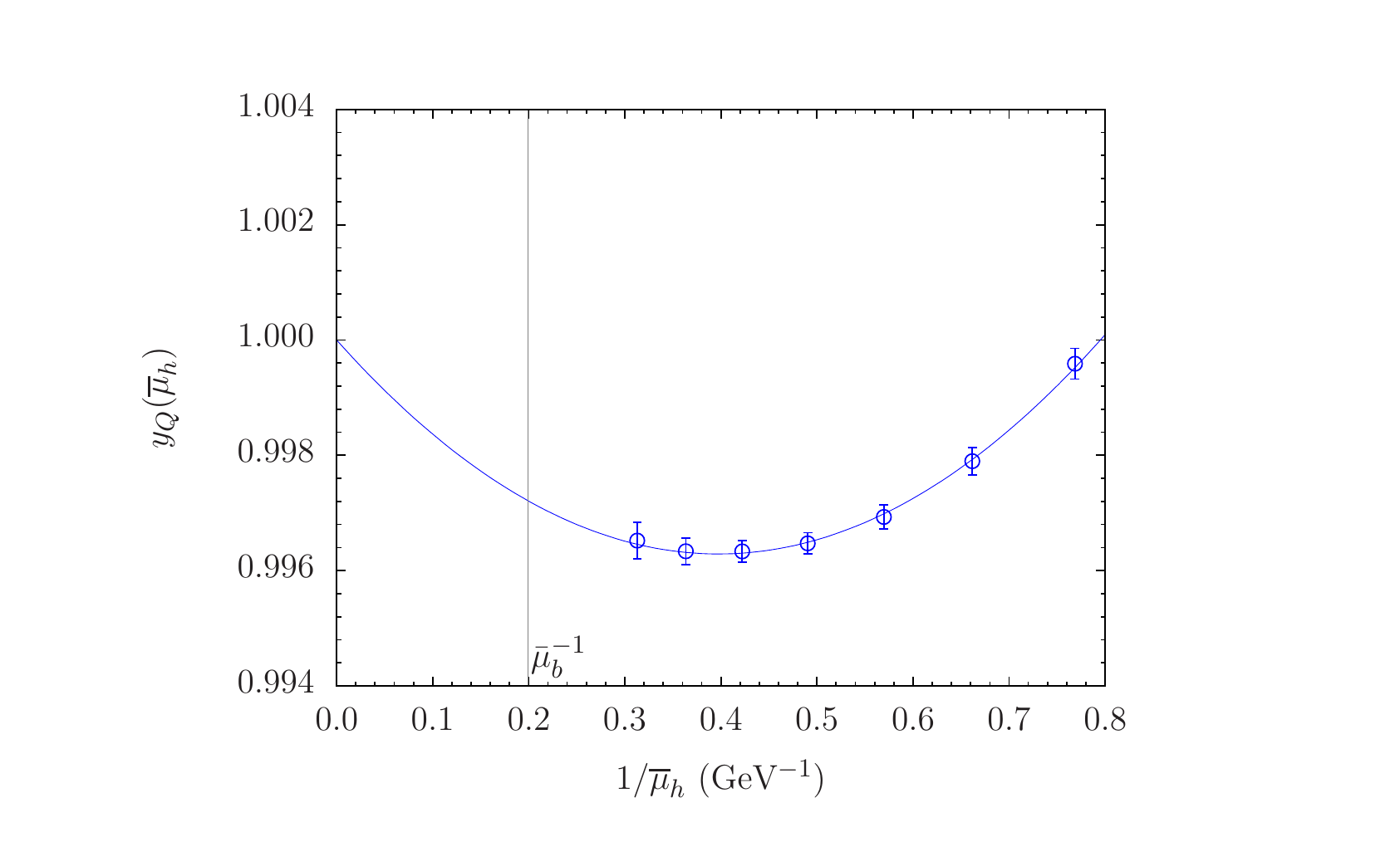}}
\caption{  $y_{Q}(\overline{\mu}_h)$ against $1/\overline{\mu}_h$ using the fit ansatz of Eq.~(\ref{eq:y_vs_muh}) with $\chi^2/(dof)=0.1$. 
We have used $\gamma=0.75$, $\lambda = 1.1588$ and $\Lambda_{QCD}^{N_f=4}= 296 (15)$ MeV for the running coupling  entering 
in the $\rho(\overline{\mu}_h, \mu)$ function. The vertical black thin line marks the position of $1/\overline{\mu}_b$. 
Quark mass values, $\overline{\mu}_h,~ 
\overline{\mu}_b$ are expressed in the $\overline{{\rm MS}}$-scheme at the scale of 2 GeV. 
  }
\label{fig:y_vs_muh}
\end{figure}

We perform the determination of the $b$-quark mass employing the {\it ratio method} described in detail 
in Refs.~\cite{Carrasco:2013zta, 
Blossier:2009hg, Dimopoulos:2011gx}. We present here a variant of this method and we build 
the quantity $Q_h = M_{hs}/(M_{h\ell})^{\gamma}$, where $M_{hs}$ and $M_{h\ell}$ are the heavy-strange and 
heavy-light pseudoscalar masses, respectively. The parameter $\gamma$ is a free one and may take values at will in the interval $[0,1)$. 
By HQET arguments we know that for the asymptotic behaviour we get:
\begin{equation}
\displaystyle \lim_{\mu_h^{\rm pole}\to \infty}  \left(\frac{M_{hs}/(M_{h\ell})^{\gamma}}{(\mu_h^{\rm pole})^{(1-\gamma)}}\right) 
= {\rm const.} ~, 
\end{equation}    
where $\mu_h^{\rm pole}$ is the heavy quark pole mass. We then consider a sequence of heavy quark masses expressed in the
$\overline{{\rm MS}}$-scheme at the scale of 2 GeV such that any two successive masses have a common and fixed ratio i.e. 
$\overline{\mu}^{(n)} = \lambda \overline{\mu}^{(n-1)}$, $n = 2, 3,  \dots $. The next step is to construct at each value of the sea 
quark mass and the lattice spacing the following ratios:
\begin{eqnarray}
&& y_Q(\overline\mu_h^{(n)},\lambda;\overline\mu_{\ell}, \overline\mu_s, a) \equiv \nonumber \\
&&\equiv 
\frac{Q_h(\overline\mu_h^{(n)};\overline\mu_\ell, \overline\mu_s, a)}{Q_h(\overline\mu_h^{(n-1)};\overline\mu_\ell, \overline\mu_s, a)} 
\cdot
\left(\frac{\overline\mu^{(n)}_h \rho (\overline\mu^{(n)}_h,\mu^{*})}
{\overline\mu^{(n-1)}_h \rho (\overline\mu^{(n-1)}_h,\mu^{*})}\right)^{(\gamma - 1) } \nonumber \\
\hspace*{-1cm}&& = \lambda^{(\gamma-1)} \frac{Q_h(\overline\mu_h^{(n)};\overline\mu_\ell, \overline\mu_s, a)}
{Q_h(\overline\mu^{(n)}_h/\lambda;\overline\mu_\ell, \overline\mu_s, a)}
\left(\frac{ \rho (\overline\mu^{(n)}_h,\mu^{*} )}{\rho( \overline\mu^{(n)}_h/\lambda,\mu^{*})}\right)^{(\gamma - 1)}\,  \label{eq:yn}
\end{eqnarray}    
with $n = 2, 3,  \dots $ and we have used the relation $\mu_{h}^{pole} = \rho( \overline\mu_h,\mu^{*})~\overline{\mu}_h(\mu^{*})$  
between the $\overline{{\rm MS}}$ renormalised quark mass (at the scale of 2 GeV) and the pole quark mass.
The $\rho$'s are known perturbatively up to N$^3$LO. For each pair of heavy quark masses we then carry out a simultaneous chiral and 
continuum fit of the quantity defined in Eq.~(\ref{eq:yn}) to obtain
$y_Q(\overline\mu_h) \equiv y_Q(\overline\mu_h,\lambda;\overline\mu_{u/d}, \overline\mu_s, a=0)$. 
By construction this quantity involves (double) ratios of 
pseudoscalar meson masses at successive values of the heavy quark mass, so  we expect that discretisation errors will be under control. 
In fact this is the case even for the largest values of the heavy quark mass used in this work, see Fig.~\ref{fig:y8}. 
Since we have taken into account the matching of QCD onto HQET concerning the evaluation of a heavy-light pseudoscalar mass,  
$M_{hs/\ell}$, our ratio $y_Q(\overline\mu_h)$ has been defined in such a way that the following ansatz is sufficient to describe the 
$\overline\mu_h$-dependence\footnote{For more details on this point see the Appendix of Ref.~\cite{Dimopoulos:2011gx} 
and \cite{Carrasco:2013zta}, section 4.} 
\begin{equation}\label{eq:y_vs_muh}
y_Q(\overline\mu_h) = 1 + \frac{\eta_1}{\overline \mu_h} +  \frac{\eta_2}{{\overline \mu}_h^2}, 
\end{equation}  
in which the constraint $\lim_{\overline\mu_h \to \infty} y_Q(\overline\mu_h) = 1$ has already been incorporated. This fit is reported in 
Fig.~\ref{fig:y_vs_muh}. 
Finally, we compute the $b$-quark mass via the {\it chain} equation 
\begin{eqnarray}
&&y_Q(\overline\mu_h^{(2)})\, y_Q(\overline\mu_h^{(3)})\,\ldots \, y_Q(\overline\mu_h^{(K+1)})= \nonumber \\
&& =\displaystyle \lambda^{K(\gamma - 1)} \,
\frac{Q_{h}(\overline\mu_h^{(K+1)})}{Q_{h}(\overline\mu_h^{(1)})} \cdot
\Big{(}\frac{\rho( \overline\mu_h^{(K+1)},\mu^*)}{\rho( \overline\mu_h^{(1)},\mu^*)}\Big{)}^{\gamma - 1} \label{eq:chain}
\end{eqnarray}
in which the values of the factors in the (lhs) are evaluated using the result of the fit function (Eq.~\ref{eq:y_vs_muh}) 
and $\lambda$, $K$ 
(integer) and $\overline{\mu}^{(1)}_h$
are such that  the quantity $Q_{h}(\overline\mu_h^{(K+1)})$  matches $M_{Bs}/(M_B)^{\gamma}$, where $M_{Bs} = 5366.7(4)$ MeV 
and $M_B$ = 5279.3(3) MeV are the experimental values of the $B_s$- and $B$-meson masses~\cite{Agashe:2014kda}, respectively. 
Notice that the quantity 
$Q_{h}(\overline\mu_h^{(1)})$ for {\it any} value of  $\overline\mu_h^{(1)}$ around the charm quark mass 
is safely computed in the continuum limit and 
at the physical pion mass. For instance, using quark mass RC of the M2-type (see \cite{Carrasco:2014cwa}) and setting as input 
$\overline{\mu}^{(1)}_h = 1.148$ GeV and $\gamma = 0.75$ we find $(\lambda, ~ K) = (1.1588, ~10)$. Thus, the $b$-quark mass in the 
 $\overline{{\rm MS}}$-scheme at the scale of 2 GeV is given by 
 $\overline{\mu}_b = \lambda^{K}~ \overline{\mu}^{(1)}_h $. Our {\it preliminary} result for the $b$-quark mass 
is given by the average over two estimates obtained using either M1 or M2-type quark mass RCs while their half difference is taken 
as an additional systematic error. This reads
\begin{equation}
m_b(\overline{{\rm MS}}, m_b)  = 4.26 (7)_{stat+fit} (14)_{syst} [16]~ {\rm  GeV}, 
\end{equation}    
where the total error (in brackets) is the sum in quadrature of the statistical and the systematic ones. For a complete 
error budget we refer to Table~3. 
\begin{table}[!h] \label{Tab: budget_mb}
\begin{center}
%\footnotesize
\scalebox{0.90}{
\begin{tabular}{|l|c|c|}
\hline
uncertainty (in \%)  & $m_{b}$ & $m_{b}/m_{c}$\tabularnewline
\hline
\hline
stat+fit & 1.6 & 1.4\tabularnewline
\hline
syst. from lat. scale & 2.6 & -\tabularnewline
\hline
syst. from discr. effects & 0.7 & 0.7\tabularnewline
\hline
syst. from ratios fits & 1.1 & 0.7\tabularnewline
\hline
syst. from chiral fits & 0.4 & 0.4\tabularnewline
\hline
syst. from RC & 1.4 & -\tabularnewline
\hline
Total & 3.6 & 1.8\tabularnewline
\hline
\end{tabular}
}
\vspace*{-0.15cm}
\caption{Full error budget for $m_b$ and $m_b/m_c$. The different sources of uncertainty are self explanatory. } 
\end{center}
\end{table}
We have verified that for a large range of values of $\gamma \in [0,~ 1)$ 
we get fully compatible final results\footnote{This systematic uncertainty has been 
included in the estimate called "syst. from ratios fits" of  Table~3.} for $m_b$. 
The freedom of choosing $\gamma$ allows for better control of systematic uncertainties stemming from discretisation 
effects and the fit ansatz Eq.~(\ref{eq:y_vs_muh}). 

\begin{figure}[!t]
\begin{center}
{\includegraphics[scale=0.55,angle=-0]{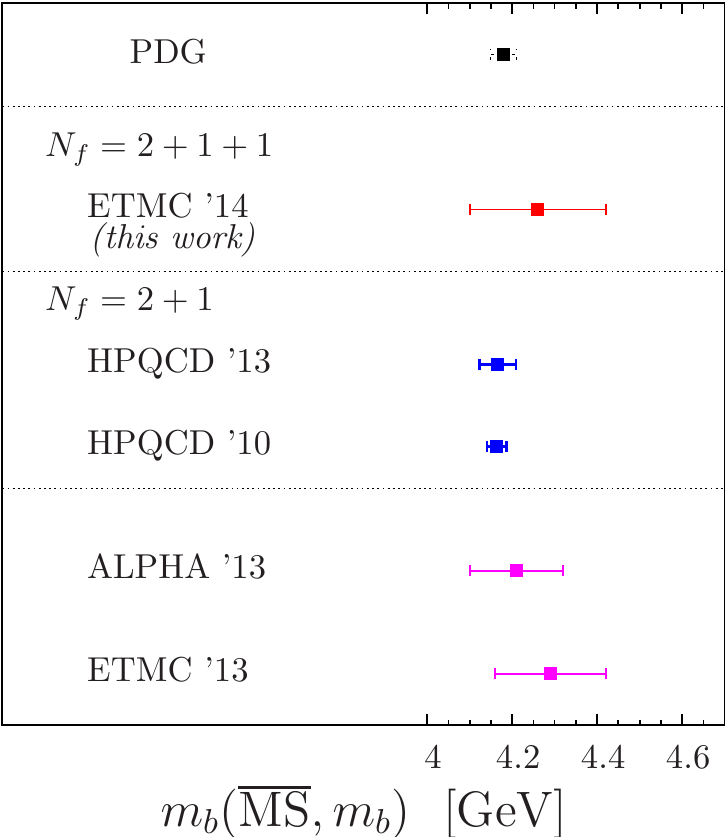}}
\caption{ We compare the available continuum limit determinations for $m_b(m_b)$ (in GeV) from lattice studies with 
$N_f=2$, 2+1 and 2+1+1 dynamical flavours. "ETMC '14" result refers to the present work. 
For the results of  other lattice studies we refer to (from top to bottom)~\cite{Lee:2013mla, McNeile:2010ji, 
Bernardoni:2013xba, Carrasco:2013zta}. For the PDG 
value see~\cite{Agashe:2014kda}. 
  }
\label{fig:mb_compr}
\end{center}
\end{figure}

\begin{figure}[!t]
\begin{center}
{\includegraphics[scale=0.55,angle=-0]{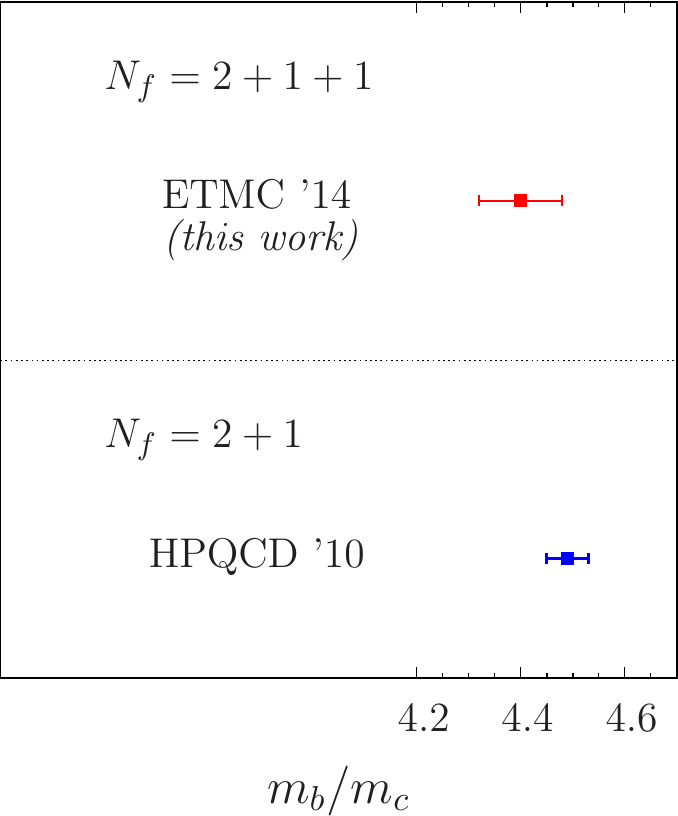}}
\caption{ Comparison between the two available continuum limit determinations for $m_b/m_c$  obtained from fully non-perturbative studies. 
For the ``HPQCD" result see~\cite{McNeile:2010ji}. 
  }
\label{fig:mb_ov_mc_compr}
\end{center}
\end{figure}

Finally, the ratio method offers the advantage of determining the ratio $m_b/m_c$ in a simple and fully non-perturbative way. By setting  
$\overline{\mu}^{(1)}_h = \overline{\mu}_c$ we repeat the above ratio method analysis and we find 
\begin{equation}
m_b / m_c  = 4.40 (6)_{stat+fit} (5)_{syst} [8]
\end{equation} 
A complete error budget is also reported in Table~3. In Figs.~\ref{fig:mb_compr} and \ref{fig:mb_ov_mc_compr} we present a  comparison 
between lattice results for $m_b$ and $m_b/m_c$, respectively. For non-lattice estimate of $m_b$ see~\cite{Chetyrkin:2009fv}.

\vspace*{0.2cm}
{\bf Acknowledgements} \\
We are grateful to all members of ETMC for fruitful discussions.  We acknowledge the CPU time provided 
by the PRACE Research Infrastructure under the project PRA067 at the J\"ulich  and CINECA SuperComputing Centers,
and by the agreement between INFN and CINECA under the specific initiative INFN-lqcd123.

%\nocite{*}
\bibliographystyle{elsarticle-num}
%\bibliography{martin}
\bibliography{biblio}





% \end{thebibliography}

\end{document}